\documentclass[%
 aip,
rsi,%
 amsmath,amssymb,
 reprint,%
]{revtex4-1}

\usepackage{booktabs}

\usepackage[dvipsnames]{xcolor}
\usepackage{graphicx}
\usepackage[normalem]{ulem}

\usepackage{dcolumn}
\usepackage{bm}

\usepackage[colorlinks=true, allcolors=blue]{hyperref}

\begin{document}

\preprint{AIP/123-QED}

\title[]{Relativistic Plasma Physics in Supercritical Fields}
\thanks{Invited Perspective Article}

\author{P. Zhang}
\email{pz@egr.msu.edu}
 \affiliation{ 
Department of Electrical and Computer Engineering, Michigan State University, East Lansing, Michigan 48824-1226, USA
}%
\author{S. S. Bulanov}
\email{sbulanov@lbl.gov}
\affiliation{Lawrence Berkeley National Laboratory, Berkeley, California 94720, USA
}%
\author{D. Seipt}
\email{dseipt@umich.edu}
\affiliation{Center for Ultrafast Optical Science, University of Michigan, Ann Arbor, Michigan 481099-2099, USA}

\author{A. V. Arefiev}
\email{aarefiev@eng.ucsd.edu}
\affiliation{Department of Mechanical and Aerospace Engineering, University of California at San Diego, La Jolla, California 92093, USA
}%
\author{A. G. R. Thomas}%
\email{agrt@umich.edu}
\affiliation{Center for Ultrafast Optical Science, University of Michigan, Ann Arbor, Michigan 481099-2099, USA}

\date{\today}

\begin{abstract}
Since the invention of chirped pulse amplification, which was recognized by a Nobel prize in physics in 2018, there has been a continuing increase in available laser intensity. Combined with advances in our understanding of the kinetics of relativistic plasma, studies of laser-plasma interactions  are entering a new regime where the physics of relativistic plasmas is strongly affected by strong-field quantum electrodynamics (QED) processes, including hard photon emission and electron-positron ($e^+$-$e^-$) pair production. This coupling of quantum emission processes and relativistic collective particle dynamics can result in dramatically new plasma physics phenomena, such as the generation of dense $e^+$-$e^-$ pair plasma from near vacuum, complete laser energy absorption by QED processes or the stopping of an ultrarelativistic electron beam, which could penetrate a cm of lead, by a hair's breadth of laser light. In addition to being of fundamental interest, it is crucial to study this new regime to understand the next generation of ultra-high intensity laser-matter experiments and their resulting applications, such as high energy ion, electron, positron, and photon sources for fundamental physics studies, medical radiotherapy, and  next generation radiography for homeland security and industry. 
\end{abstract}

\maketitle

\section{\label{sec:level1}Introduction}
One of the predictions of quantum electrodynamics (QED) is that in the presence of an electric field stronger than a critical field strength, `break-down' of the vacuum occurs, which results in the spontaneous creation of matter and antimatter in the form of electrons and positrons \cite{Sauter1931,Heisenberg_Zphys_1936,Schwinger_PR_1951}. Extremely strong fields can also polarize the quantum vacuum; predicted exotic effects include light-by-light scattering, vacuum birefringence, 4-wave mixing, or high order harmonics generation from the vacuum \cite{Toll_Thesis,MTB_RMP_2006,Heinzl:OptCommun2006,Marklund_RMP_2006,DiPiazza_RMP_2012,Piazza-PRD-2005,King_JP_2016,Fedotov_PLA_2007}. While QED is probably one of the best verified theories so far on a single particle level \cite{Gabrielse2006,PhysRevD.98.030001}, the new collective phenomena that arise when electrons, positrons, and photons are exposed to \textit{strong} electromagnetic  fields are not yet well understood. Strong electric fields are those that approach or exceed the \textit{QED critical field strength}, $E_{cr}$, in which interactions become highly nonlinear. In particular, the prolific production of electrons and positrons can cause complex plasma interactions with these fields, which are conditions that are only starting to be theoretically explored. The physics of such plasmas in strong fields is relevant to early universe conditions\cite{QED_strong_fields_book}, extreme astrophysical objects such as neutron star atmospheres \cite{Goldreich_PRL_1969} and black hole environments \cite{Ruffini_PR_2010}, and is critical to future high-intensity laser driven relativistic plasma physics. 

Strong field (SF) QED processes have for a long time been believed to be the domain of elementary particle physics \cite{Ritus_1979,Burke_PRL_1997}, which has made tremendous progress in the last hundred years, from  the formulation of basic laws and the construction of the first particle accelerators to the creation of the elaborate Standard Model and its experimental verification at grand-scale experimental facilities, such as LEP, LEP II, SLAC, Tevatron, and LHC. One of the directions of research that follows this success is the study of the cooperative behavior of  relativistic quantum systems and the basic properties of the quantum vacuum. High power laser-plasma interactions in regimes that were not available previously can provide the framework for the study of these effects (Fig. \ref{fig:fig1}).

The highest laser intensities demonstrated to date are seven
orders of magnitude lower than necessary to reach the critical field $E_{cr}$. However, since the electric field is not a Lorentz invariant, in the rest frame of an ultrarelativistic particle a subcritical field strength may be boosted to the critical field strength and beyond. Therefore, SF QED processes such as multiphoton Compton emission of photons and multiphoton Breit-Wheeler electron-positron pair production occur at significantly lower field strengths than $E_{cr}$. This allows studies of the \textit{physics of plasmas in supercritical fields} with present day and near future technology. Particle accelerators or extremely powerful lasers are able to generate high-energy particles that can experience these boosted field strengths (e.g. return forces from ions in plasmas, or fields in standing waves).  

Laser fields may provide both the strong electromagnetic field and generate the high-energy particles \cite{Bell_PRL_2008} and therefore represent a particularly interesting environment for studying plasma physics in supercritically strong fields \cite{MTB_RMP_2006,Marklund_RMP_2006,DiPiazza_RMP_2012}. Despite tremendous progress achieved in recent years, there are a lot of unanswered questions and unsolved problems that need to be addressed both theoretically and in experiments.  

In this perspective article, we will give an introduction to the physics of relativistic plasma in supercritical fields, discuss the current state of the field and give an overview of recent developments, and highlight open questions and topics that, in our opinion, will dominate the attention of people working in the field over the next decade or so. This is not a comprehensive review, comes from a US perspective, and the reader is referred to other papers on the broader topic of the physics of extremely high intensity lasers, such as ref.~\cite{DiPiazza_RMP_2012}. 

The structure of this perspective article is as follows. In section \ref{sec:level2a} we describe the parameter space of plasma in strong fields. Then in section \ref{sec:level2}, we give an overview of recent developments in strong field electrodynamics and discuss some examples where QED strongly affects plasma dynamics. Finally, in section \ref{sec:level7} we outline strategies for progress in this field.

These stategies lead us to envision three facility configurations to study this physics. The first one will feature a laser pulse colliding with an ultrarelativistic electron beam with energy $\mathcal{E}_\mathrm{beam}$ such that the product 
$\left(\mathcal{E}_\mathrm{beam}\;{\rm[GeV]}\right) \sqrt{P_\mathrm{laser}\;{\rm[PW]}}\left(\lambda_\mathrm{laser}\;{\rm[\mu m]}\right)^{-1}\gg1$, where $P_\mathrm{laser}$ is the laser pulse peak power of wavelength $\lambda_\mathrm{laser}$. The second one will employ multiple colliding laser pulses  that satisfy the condition  $\left(P_\mathrm{laser}\;{\rm[PW]}\right)\left(\lambda_\mathrm{laser}\;{\rm[\mu m]}\right)^{-1}\gg10$. Finally, the third one will combine the capabilities of the first two with that of the laser plasma based collider\cite{Leemans_PhysToday_2009} for the studies of plasma physics in supercritically strong fields at the highest intensities. 
To date, experiments in this research area have all been performed using the first configuration of a laser colliding with an electron beam, but have not yet reached the critical field strength (although several notable experiments have come close). With new facilities capable of achieving greater than 10 PW power, it will become possible to experimentally reach a new \textit{radiation dominated regime} \cite{Bulanov_NIMPR_2011} where phenomena, including prolific creation of matter from light, can be explored.

\section{\label{sec:level2a}Plasma in supercritical fields}
\subsection{What is a supercritical field?}
In classical electrodynamics the interaction strength of electromagnetic (EM) fields with charged particles is usually described in terms of the dimensionless amplitude of the electromagnetic vector potential, $a_0=eE/m\omega c$, which
represents the work done by the fields over a distance $\lambda/2\pi$ in units of the {electron} rest mass energy $mc^2$. Here $\omega$ and $\lambda$ are the (oscillating) electromagnetic field frequency and wavelength, $E$ is the amplitude, ${-} e$ and $m$ are the electron charge and mass, $c$ is the speed of light. Hence, when $a_0\gtrsim 1$ the interaction is always relativistic and nonlinear. Short laser pulses of that intensity ($I \gtrsim 2\times10^{18}$ W/cm$^2$ for a 800 nm laser) will drive large amplitude waves in underdense plasmas which in turn can accelerate electrons to multi-GeV energies, so-called laser-wakefield acceleration (LWFA) \cite{ESL_RMP_2009}. Laser ion acceleration from overcritical plasmas requires higher intensities of $10^{19}-10^{22}$ W/cm$^2$  ($a_0\sim 10 - 100$ for a 800 nm laser) \cite{MTB_RMP_2006,Macchi_RMP_2013,Daido_RPP_2012}, and the highest intensity demonstrated up to now is ${5.5}
\times10^{22}$ W/cm$^2$ (see Refs. {\citet{Yoon_OP_19},} \citet{Yanovsky:08,Kiriyama:18}). Next generation laser facilities \cite{Danson_2019}, such as Extreme Light Infrastructure (ELI) \cite{eli}, APOLLON \cite{Apollon}, Center for relativistic Laser Science (CORELS) \cite{CORELS}, EP OPAL \cite{Bromage_2019}, Zetawatt Equivalent Ultrashort pulse laser System (ZEUS) and Station of Extreme Light (SEL) are planned to be able to achieve $10^{23}-10^{24}$ W/cm$^2$. 

In addition, $a_0$ also controls multi-photon interactions in QED scattering processes \cite{Ritus_1985}; the number of laser photons in a cylindrical volume of length $\lambda/2\pi$ and transverse dimension of the (reduced) Compton length $\lambdabar_C = \hbar /mc$ is $a_0^2 /(4\pi\alpha)$, with fine structure constant $\alpha$. Thus, $a_0 {> } 1$ as the effective interaction strength signals {that the interaction might have entered the multi-photon regime.}

Strong field QED describes the physics of particles in such strong EM
fields and is broadly characterized by environments where the strength of electric fields is 
 of the order of the QED critical field 
\begin{align}
E_{cr} = \frac{m_e^2c^3}{e\hbar} = 1.3\times 10^{18}\, \mathrm{V/m} \,, 
\end{align}
which is the field that classically would accelerate an electron to its rest mass energy in a Compton length \cite{Schwinger_PR_1951}. Vacuum pair production, known as the Sauter-Schwinger process, can be understood as virtual dipole pairs in the vacuum being torn apart by the field, becoming real asymptotic pairs, the probability of which scales as $\propto \exp \{- \pi E_{cr}/E\}$ in a constant $E$ field \cite{Schwinger_PR_1951}. Such a field strength corresponds to a light intensity of $4.65\times 10^{29}$ W/cm$^2$ and a dimensionless amplitude, $a_S=mc^2/\hbar \omega$, of $4.1\times 10^5$ for a 1 $\mu$m laser field. At fields of this strength, QED processes are highly nonlinear and cannot be described by straightforward perturbation theories. 

The relevant parameter that characterizes the interaction of electrons, positrons, and photons with strong EM fields is 
\begin{equation}
\chi = \frac{|F^{\mu\nu}p_\nu|}{ m E_{cr} } \,,
\end{equation}
where $F^{\mu\nu}$ is the EM field tensor, and $p_{\nu}$ is the corresponding particle 4-momentum. For electrons and positrons the parameter $\chi$ has a simple physical meaning, {\it i.e.}, the EM field strength in the rest frame of the particle. Hence, for relativistic plasma where the bulk of the particles experience $\chi\gtrsim 1$, we define the fields to be \textit{supercritical}.

\begin{figure*}
\begin{center}
    \includegraphics[width=0.6\textwidth]{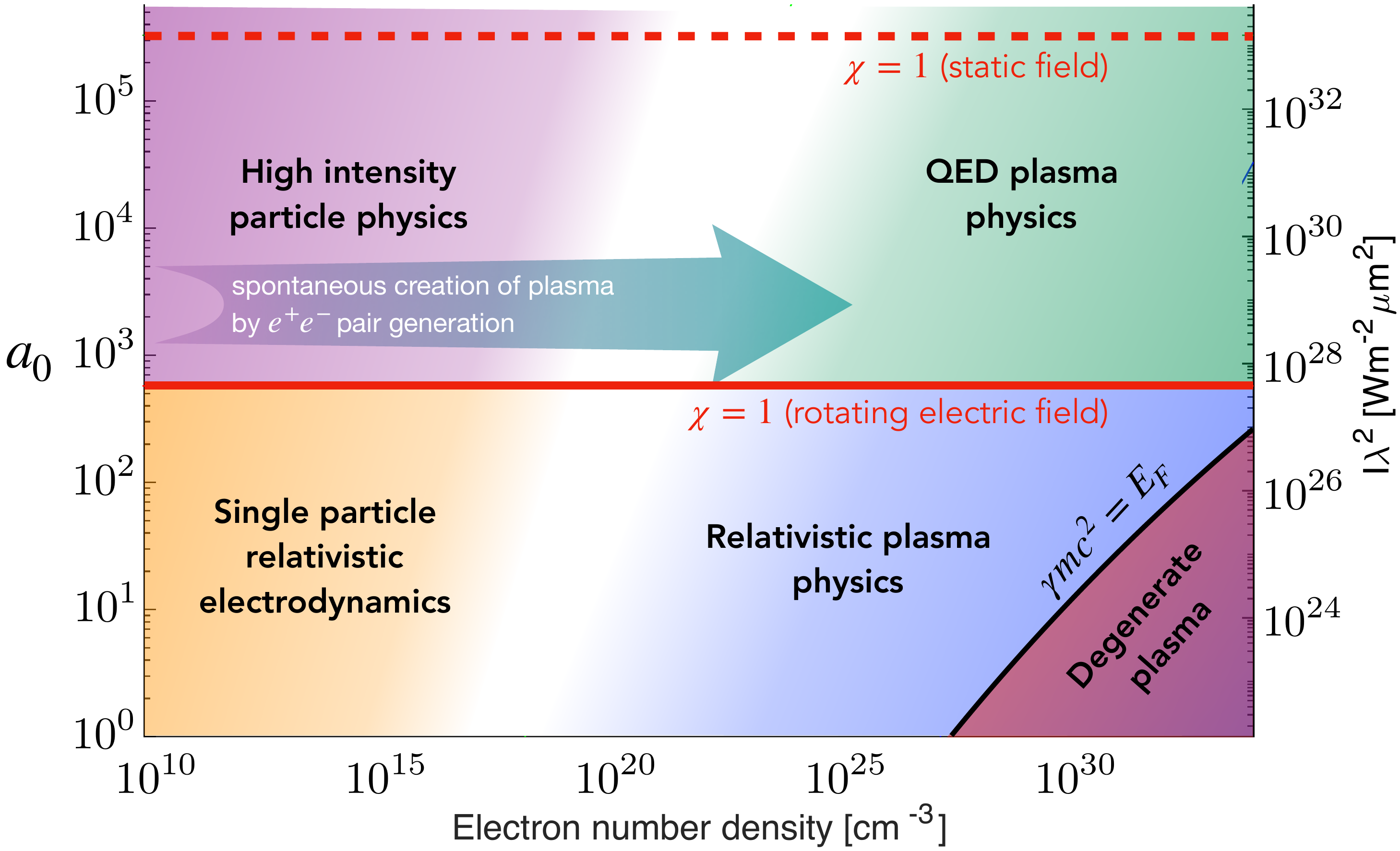} 
\caption{Different regimes of strong field physics as a function of plasma density and either field strength $a_0$ (left scale) or laser intensity (right scales). The line $\chi = 1$ in the electron rest frame assumes a rotating configuration with the electron Lorentz factor being $\gamma = a_0$.}
\label{fig:fig1}
\end{center}
\end{figure*}

There are two primary ways to achieve high $\chi$ for electron-EM field interaction. By assuming a plane EM 
wave ($|\mathbf{E}|=c|\mathbf{B}|$ and $\mathbf{E}\perp \mathbf{B}$), we have $\chi=(\gamma |E|/E_{cr})(1-\beta\cos{\theta_k})$, where $\gamma$ is the Lorentz factor, $\beta$ is the electron velocity, and $\theta_k$ is the angle between the electron momentum direction and the wave vector. Thus, $\chi$ can be maximized for electrons counter-propagating with respect to wave vector ($\theta_k=\pi$), i.e. an electron beam colliding with a laser field. On the other hand, assuming $|\mathbf{B}|=0$ gives $\chi=(\gamma \hbar|E|/E_{cr})\sqrt{1-\beta^2\cos^2{\theta_e}}$, where $\theta_e$ is the angle between the electron momentum and the electric field. In this case, $\chi$ is maximized when electron momentum is perpendicular to the electric field ($\theta_e=\pi/2$), which can be achieved with colliding laser pulses (at the magnetic nodes, where only rotating electric field is present). {In such a circular polarization, the electric field and electron momentum in equilibrium motion (i.e. circulating) are (close to) perpendicular. In a linearly polarized case, however, they are parallel, so a boost from the laser electric field cannot be obtained directly. It is possible to have an electron with momentum perpendicular to the linearly polarized field, but this has to be externally supplied (with an accelerator).} Apart from maximizing $\chi$ these two configurations represent two fundamental schemes for the study of strong field (SF)
QED phenomena. The interaction of a high energy electron with a plane EM wave can be considered as a one-dimensional problem, when the energy of an electron is high enough to neglect the transverse effect of the EM field. In this case the dynamics of an electron is fully determined by its energy dissipation due to the radiation reaction effects, and permits an analytical solution {\cite{Landau_1975, Piazza_LMP_2008}}. In the second case, the electron dynamics is determined by the transverse motion in the rotating electric field and, also, permits an analytical solution. The rotating field configuration can be achieved in practice with colliding laser fields or a laser reflecting from a dense surface \cite{Bell_PRL_2008}.

\subsection{Basic QED processes in strong fields}

\begin{figure}[b]
	\begin{center}
		\includegraphics[width=0.4\textwidth]{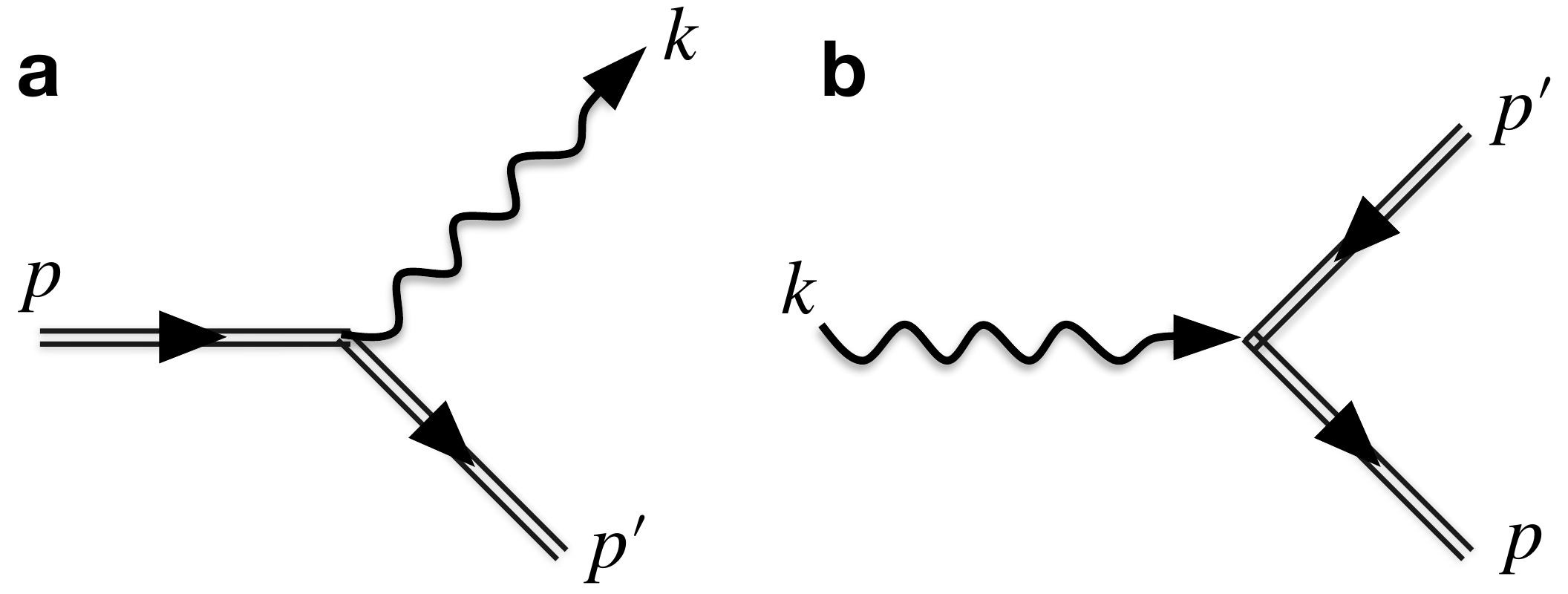}
		\caption{The two lowest order processes in strong field QED. (a) emission of a photon by a \emph{dressed} lepton. (b) decay of a photon into a \emph{dressed} electron-positron pair. {Double lines denote Volkov electrons \cite{Wolkow_ZPhys_1935}, i.e. electrons
				dressed through the interaction with the laser photons.
		}}
		\label{fig:fig1p5}
	\end{center}
\end{figure}

The two lowest order processes in strong field QED theory are  illustrated in Fig. \ref{fig:fig1p5}. These are (a) emission of a photon by a \emph{dressed} lepton and (b) decay of a photon into a \emph{dressed} electron-positron pair. A dressed lepton means it is a particle state in a background electromagnetic field and not a free (vacuum) particle state as is usual. In a plane wave, the expectation of the momentum of a dressed lepton will be its canonical momentum in classical mechanics. These two processes are known as \emph{multiphoton Compton} emission (in a plane wave, or \emph{synchrotron emission/magnetic bremsstrahlung} in a constant field) and \emph{multiphoton} Breit-Wheeler pair-production, respectively. As the lowest order processes, these  dominate high order processes by a factor $\sim\alpha = 1/137$ in probability and thus are the most important to consider for their effect on plasma dynamics. {Other processes of the same order in $\alpha$, such as annihilation processes, are negligible because of phase space suppression.}

When the classical field strength parameter $a_0$ is large, the coherence length (or ``formation length'') of any quantum process is short and therefore it is typical to assume that the emission processes can be approximated by those in constant fields, the so called Local Constant Field Approximation (LCFA).  This is usually the case if\cite{Dinu_PRL2016,DiPiazza_PRA_2019} $1 \ll a_0 $ and $\chi \ll a_0^3$. 
The formation length being short is one of the basis assumptions for standard simulation codes, as described later. {Typically, for a plane wave the formation length is estimated as $L_f=1/a_0\omega$ \cite{Ritus_1979}. However, the formation length depends on the emitted photon energy, $L_f=L_f(\omega^\prime)$, and for sufficiently low $\omega^\prime$ the formation length can become of the order of a laser wavelength, making the LCFA invalid \cite{Piazza_PRA_2018,Blackburn_beaming}.}

These two quantum processes affect the dynamics of the \emph{plasma} by modifying the
\begin{itemize}
\item electron kinetics through momentum loss in energetic photon emission (``radiation reaction'').
\item plasma density through electron-positron pair production.
\end{itemize}
These quantum effects would therefore continuously change the basic plasma parameters (e.g. plasma density, plasma temperature, and plasma frequency etc) during the interaction of light and matter. As a result, the collective behavior of QED plasmas would be very different from those of the classical plasmas. 

\begin{figure}
\begin{center}
    \includegraphics[width=0.5\textwidth]{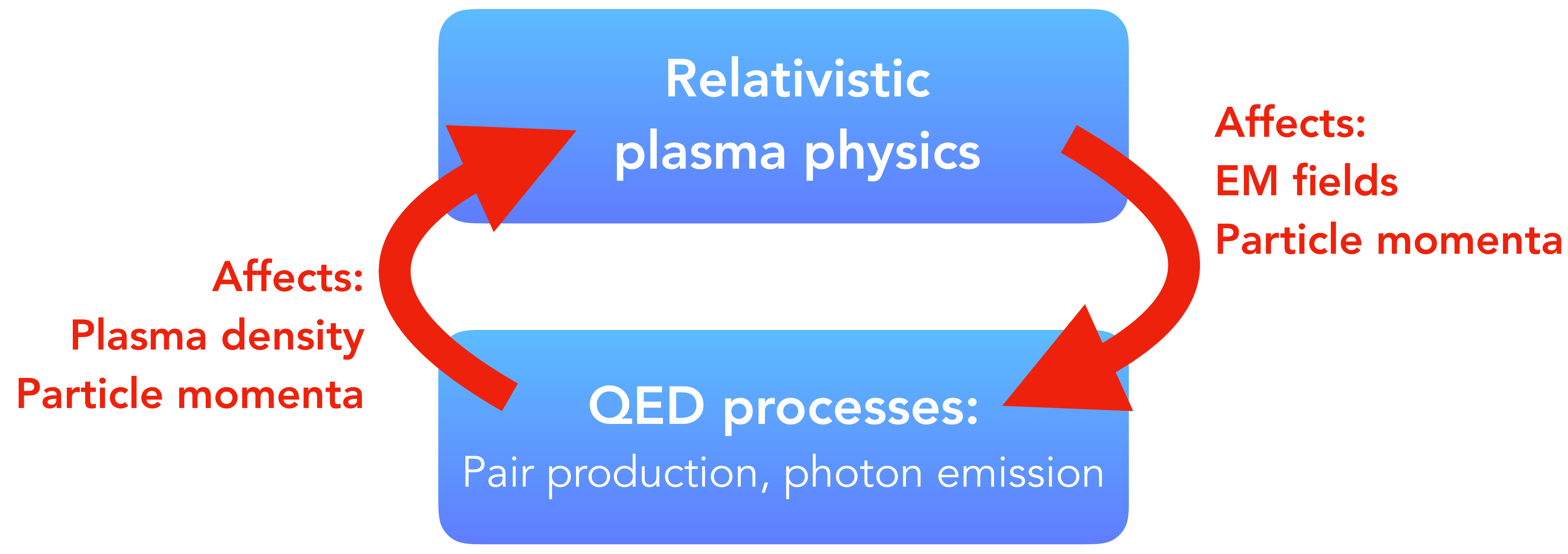}
\caption{The coupling of QED processes and relativistic plasma dynamics.}
\label{fig:fig1p75}
\end{center}
\end{figure}

Moreover, since the quantum processes themselves depend on the momentum distribution of the particles and the EM fields generated by the plasma dynamics, there is a strong coupling between the two which should lead to rich new phenomena.  
The complex feedback between QED and collective plasma processes (Fig. \ref{fig:fig1p75}) is what makes the QED-plasma regime unique .

\subsection{Collective effects in supercritical fields}

The connection of these strong field QED processes to plasma physics also requires consideration of collective interactions between particles. 
To illustrate this we sketch the landscape of laser-plasma interactions in the (plasma density, laser intensity) plane in Fig.~\ref{fig:fig1}, with the underlying assumption of a rotating field configuration. At lower laser intensities and particle densities, the particle trajectories are determined by their classical dynamics in the laser field alone without the influence of collective effects, i.e., it is ``Single particle electrodynamics''. As the density of particles increases, collective plasma effects start to dominate the single particle dynamics. {The boundary between collective and single particle motion is defined using the threshold where the Coulomb force due to the fully perturbed ($\delta{n}/n\sim1$) plasma balances the ponderomotive force due to the laser, $a_0=4\pi{r_e}n_e/k_0^2$, where $n_e$ is the electron number density.} This is when the interaction enters the domain of ``Relativistic plasma physics'', in which interesting physics phenomena such as plasma wakefield acceleration \cite{ESL_RMP_2009} and laser driven ion acceleration \cite{MTB_RMP_2006,Macchi_RMP_2013,Daido_RPP_2012} may occur. Even higher particle densities result in particle kinetic energies becoming equivalent to their Fermi energy, which is characteristic of the ``Degenerate Plasma'' regime. {The Fermi energy for relativistic plasma is $E_F=\sqrt{\hbar^2(3\pi^{2}n_e)^{2/3}c^2+m_e^2c^4}-m_e{c^2}$.
}

At higher laser intensities but low particle densities, the interactions are in the domain of "High intensity particle physics". Here, the particle dynamics is dominated by radiation emission and quantum processes including interactions with the quantum vacuum, but collective effects are negligible. For example, light-by-light scattering, thought to be responsible for the attenuation of X-rays by background light in cosmology, is such a process. Under certain conditions, the spontaneous generation of electron, positron, and photon plasma in strong fields becomes possible, which is usually referred to as EM cascades (shower or avalanche type, see discussion below in Sec. \ref{sec:IIIb}). This prolific plasma creation in high-intensity laser fields rapidly pushes the interaction into the ``QED plasma physics'' domain, where both collective and quantum processes determine the particle dynamics. Production of dense electron, positron, and photon plasma will provide new opportunities for laboratory studies of the most extreme astrophysical environments.

There are two natural QED thresholds in this picture, the QED critical field in the laboratory frame (dotted red line) and the QED critical field in the particle rest frame (solid red line). Experimental results achieved up to date all lie below the dotted line in 
Fig. 1, including demonstrations of matter creation from light \cite{Burke_PRL_1997} and quantum radiation reaction \cite{Cole_PRX_2018,Poder_PRX_2018}. Theoretical research studies have only recently started to explore physics beyond this boundary. Already at this threshold, the particle dynamics is dominated by radiation emission and is not completely understood because of the approximations required in the theory to obtain tractable solutions. Hence, achieving super-critical fields in plasma is a frontier area of research. 
\subsection{Connections to Astrophysics}

QED plasma is of interest in many fields of physics and astrophysics. The new plasma state that is created in the presence of supercritical fields is similar
to that thought to exist in extreme astrophysical environments including the magnetospheres of pulsars and active black holes. Electron-positron plasmas are a prominent feature of the winds from pulsars \cite{Goldreich_PRL_1969} and black holes \cite{Ruffini_PR_2010}. {In these environments where the fields are typically purely magnetic and particles are typically ultrarelativistic, one can use the crossed-field configuration in place of a magnetic field. This is because $\mathbf{E}\cdot\mathbf{B}$ is a Lorentz invariant, and the angle between $\mathbf{B'}$ and $\mathbf{E'}$ in the boosted frame is set by $\mathbf{B'}\cdot\mathbf{E'}/|\mathbf{B'}||\mathbf{E'}|=\mathbf{B}\cdot\mathbf{E}/|\mathbf{B'}||\mathbf{E'}|$. Further, $E^2-B^2$ is an invariant so $|\mathbf{B'}||\mathbf{E'}|=|\mathbf{E'}|^2\sqrt{ 1 +  (B^2-E^2)/{E'}^2}$. Hence, if the boosted electric field $|\mathbf{E'}| \gg \mathbf{|E|},\mathbf{|B|}$, then the angle between $\mathbf{E'}$ and $\mathbf{B'}$ in the boosted frame is $\cos\theta \approx (\mathbf{E}\cdot\mathbf{B})/ |\mathbf{E'}|^2 \ll 1$. Hence, ultrarelativistic particles see in their rest frame an arbitrary field as approximately crossed}.

The recent release of the first image of a black hole \cite{EHTC_2019} is expected to inspire new interest in the study of relativistic electron-positron plasma physics. The 1.3 mm wavelength image reveals an asymmetry in brightness in the ring, which is explained in terms of relativistic beaming of the emission from a plasma rotating close to the speed of light around a black hole. Some of the models suggest that the mm emission is dominated by electron-positron pairs within the funnel, even close to the horizon scale \cite{EHTC1_2019,Hirotani_APJ_2016}. Electron-positron pairs are produced from the background radiation field or from a pair cascade process following particle acceleration by unscreened electric fields. These processes would efficiently emit gamma-rays via curvature and inverse-Compton processes\cite{Hirotani_APJ_2016}. The suppression of emission from the disk and funnel wall, and the simultaneous production of a sufficiently powerful jet would be subjects of future research using pair plasma models \cite{EHTC1_2019}. The study of relativistic plasmas in supercritical fields in the laboratory may help us better understand this and  other extreme astrophysical events, such as Gamma Ray Bursts\cite{Accetta_PLB}. 
\section{\label{sec:level2}Strong Field Quantum Electrodynamics: Recent developments}

While the Sauter-Schwinger process (electron-positron pair production in vacuum) is inaccessible by present day laser and accelerator technologies, the multiphoton Compton and Breit-Wheeler processes have already been observed in experiments \cite{Burke_PRL_1997,Poder_PRX_2018,Cole_PRX_2018}. In principle, the scope of SF QED is much wider, and includes the study of strong field effects on elementary particle decays and the searches of the physics beyond the Standard Model \cite{Ritus_1979, DiPiazza_RMP_2012}. However, in terms of their effects on plasma interactions, these lowest order processes are dominant.

Most of the classical SF QED results were obtained by assuming a plane monochromatic wave or a constant crossed field ($|\mathbf{E}|=c|\mathbf{B}|$ and $\mathbf{E}\perp \mathbf{B}$), since these two configurations allow  analytical formulae for the probabilities of Compton and Breit-Wheeler processes \cite{Ritus_1979,Ehlotzky_RPP_2009,DiPiazza_RMP_2012} to be obtained quite easily. However, due to the pulsed nature of strong lasers, it became clear that the plane wave approximation is not able to adequately describe the physics of these processes. Moreover, in such fields multi-staged processes dominate the interaction, {\it i.e.}, the mean free path of an electron or positron with respect to the probability of radiating a photon is smaller than the characteristic size of the pulsed field. The same is true for a photon decay into an electron-positron pair. These considerations led to the study of Compton and Breit-Wheeler processes in short pulsed fields, as well as to the first steps in the study of multi-staged processes, including double Compton \cite{Seipt_PRD_2012,Mackenroth_PRL_2013} and trident \cite{Mackenroth_Trident,Dinu_Trident,King_PRD2018}. These studies are intrinsically connected with the calculation of higher order Feynman diagrams, which at certain field strength indicate the breakdown of the perturbation theory and the need for new paradigms in SF QED.

In what follows we briefly review the main recent developments in SF QED and the related plasma physics, which, from our point of view, illustrates both the direction the field is taking as well as the regions where we lack the understanding of physics and where the concentrated effort of the scientific community should be directed.

\subsection{Quantum Radiation Reaction}
When an accelerated charged particle emits radiation, it experiences an effective recoil force, which is usually referred to as the radiation reaction (RR)
or radiation friction. In classical electrodynamics it is a well known effect, first described by Lorentz-Abraham-Dirac equation \cite{Abraham_Annalen_1902,Dirac_PRSL_1938}. However, due to the fact that this equation has self-accelerating non-physical solutions, the Landau-Lifshitz (LL) prescription \cite{Landau_1975} is usually employed to describe the RR, which uses certain assumptions for the charged particle motion and the forces acting on it. We note that the exact form of the classical equation of motion with RR included is still under discussion in the scientific community, as well as how to derive them as a low energy limit of exact QED calculations (see Ref. \cite{ilderton.plb.2013} and references cited therein). However, as more energetic electron beams interacting with more intense lasers are being considered, the classical LL description ceases to be valid. This strong RR regime is characterized by a number of quantum effects, such as stochastic photon emission \cite{Blackburn_2014}, a hard cutoff in the emitted photon spectrum \cite{Burton_2014}, straggling \cite{Shen_PRL_1972}, quantum quenching \cite{Harvey_PRL_2017}, and trapping in travelling \cite{Zeldovich_USP_1975,Ji_PRL_2014} and standing \cite{Kirk_PPCF_2009,Gonoskov_PRL_2014,Jirka_PRE_2016} EM fields. The cut-off in the spectrum means that the radiated power is reduced compared to the classically predicted radiated power, by a factor $g(\chi) = P_{\rm quantum} / P_{\rm classical}$. Interestingly, even for $\chi = 0.1$, $g(\chi)$ is already $0.66$. This effect can be taken into account phenomenologically by modifying the LL equation. Other  quantum radiation reaction effects require full SF QED treatment of the underlying physics.  

\begin{figure}
\begin{center}
    \includegraphics[width=0.45\textwidth]{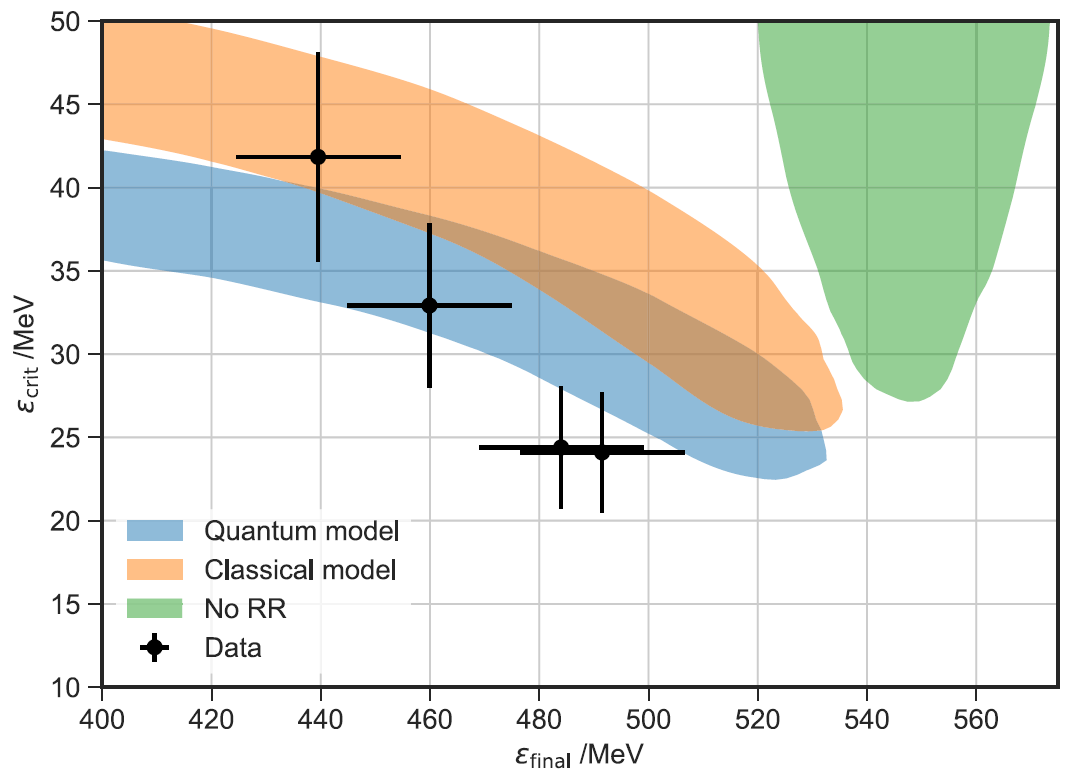}
\caption{Comparison of experimental data (points with error bars) and different models (shaded areas) for the critical energy $\epsilon_{crit}$ as a function of the postcollision energy of the electron beam $\epsilon_{final}$. $\epsilon_{crit}$ is a characteristic energy of the photon spectrum, measuring the spectral shape \cite{Cole_PRX_2018}. Reproduced with permission from J. M. Cole et al., Phys. Rev. X 8, 011020 (2018). Copyright 2018 American Physical Society.}
\label{fig:figx}
\end{center}
\end{figure}

Quantum radiation reaction has been the subject of active theoretical and computational research in the past decade (see  e.g.\cite{DiPiazza_PRL_2009,DiPiazza_PRL_2010,DiPiazza_RMP_2012,Thomas_PRX_2012,Blackburn_2014,Zhang_NJP_2015}) while the experimental effort was missing. However, this situation recently has changed with two experiments carried out on the Gemini laser \cite{Poder_PRX_2018,Cole_PRX_2018}, which were studying the interaction of GeV-class electron beams with intense laser pulses ($a_0>1$). Both experiments reported on a significant (30-40\%) electron beam energy loss after the interaction with a counter-propagating laser pulse. The analysis of the experimental data showed that the results clearly rule out a no-RR possibility (see Fig.~\ref{fig:figx}). However, these experiments were not able to accurately distinguish between the LL description and the full SF QED one, although both studies indicated that the quantum model gave better agreement. Nevertheless, they are an important milestone in the study of SF QED effects, not only because the previous study was performed at SLAC \cite{Burke_PRL_1997} more than two decades ago using conventional accelerator technology and much lower laser intensities, such that $a_0$ was small, but because at the Gemini facility laser plasma accelerated electrons{, which were of very small beam size allowing overlap between most of the electrons and the tightly focused laser beam,} were used to collide with a significantly more intense laser pulse in a high $a_0$ and high $\chi$ regime where the radiation emission plays an important role in determining the electron kinematics. {Note that quantum RR was also recently studied experimentally in aligned crystals\cite{Wistisen_NatCom2018}.} 

\subsection{Pair-plasma production}
\label{sec:IIIb}
Electron-positron pair production, as predicted by QED theory \cite{Dirac_1928}, offers the possibility of a direct transformation of light into matter, for example by the Breit-Wheeler (BW) process ($\gamma\gamma\rightarrow e^+e^-$) \cite{Breit_PR_1934}. The observation of the BW process is challenging because one needs not only high photon energies to surmount the production threshold at the rest mass energy of the pair, but also high photon densities to overcome the smallness of the cross-section and achieve an appreciable yield. Laser-driven gamma-ray sources are the key to overcoming these challenges. Two different approaches have been proposed that rely on such sources. One approach is to fire a laser generated gamma-ray beam into the high-temperature radiation field of a laser-heated hohlraum~\cite{Pike_Nature_Photonics_2014}, whereas  another proposed approach is to  collide two gamma-ray beams~\cite{Ribeyre_PhysRevE_2016, jansen2018_PPCF,Wang_PRA_2020}. Similar approaches may be extended to multiple colliding photon beams instead of two\cite{Bulanov1_PRL_2010}. Such multiple colliding photon beams may yield different scalings in terms of pair production, in which a specific $n$-photon process may be studied in isolation from other events. 

Although the (linear) BW process has not been directly observed in experiment, the conversion of a photon into electron-positron pair in the presence of a strong EM field (multiphoton BW process, $\gamma+n\gamma^\prime\rightarrow e^+e^-$) was detected at the E-144 experiment at SLAC \cite{Burke_PRL_1997}, where the 46.6 GeV electron beam from SLAC's linear accelerator collided with a counter-propagating $\sim10^{18}$W/cm$^2$ laser pulse with photon energy of 2.35eV. In the rest frame of the electrons, the laser intensity and frequency are significantly upshifted, reaching $\chi\sim 0.27$ at $a_0\sim 0.32$, to be able to generate high-energy photons from multiphoton Compton scattering, some of which were subsequently converted into $e^+e^-$ pairs in the multi-photon BW process. 

We note, the above mentioned setups do not offer the possibility of pair \emph{plasma} production, but this can be achieved even in moderate intensity laser interactions with high$-Z$ solid targets. In this case, when relativistic electrons generated in the pre-plasma propagate through the solid target, MeV bremsstrahlung photons are generated, which lead to the production of pair plasma in the field of the high-Z target nuclei through the Bethe-Heitler (BH) process \cite{Heitler_1954,Chen_PRL_2009, Sarri_PRL_2013, Chen_PRL_2015}. {Generation of neutral dense pair plasma in the laboratory is also reported\cite{Sarri_NatC15}, with observation for a Weibel like instability driven by the pair plasma \cite{Warwick_PRL17}.} When the laser intensity is increased further, prolific pair plasma production is possible, mostly due to  SF QED effects. This has been extensively studied theoretically and using particle in cell (PIC) simulations \cite{Nakamura.PRL.2012,Ridgers_PRL_2012,Zhang_NJP_2015}.

However, as the energies of particles and intensity of EM fields are increased, a new possibility for producing pair plasma arises, through a cascaded production process of electrons, positrons, and high energy photons \cite{Bell_PRL_2008,Fedotov.prl.2010,Bulanov1_PRL_2010,Sokolov_PRL_2010,Bulanov.pra.2013}. These cascades come in two types \cite{Mironov.pla.2014,Fedotov.prl.2010,Mironov.pla.2014}. The first is the shower-type cascade, where the initial particle energy is repeatedly divided between 
the products of successive Compton and BW  
processes and typically happens in the collision of a high energy particle beam with an intense laser pulse. In this case, the EM field has almost no effect on the particle trajectories. The second is the avalanche-type cascade, where the EM field both accelerates {the charged particles} and causes QED processes. In this case, the number of particles grows exponentially, fueled by the energy transformation from the EM field into electrons, positrons, and high energy photons. {It has been proposed that the maximum attainable laser intensity may be determined by the cascade development\cite{Bulanov1_PRL_2010,Fedotov.prl.2010}.}

In order to reach $\chi > 1$ with minimal total laser energy, an optimal field configuration is needed. It was first identified in the study of $e^+e^-$ pair production that colliding multiple laser pulses at one spot provides such a configuration \cite{Bulanov_PRL_2010}, which in a limiting case of many laser pulses can be seen as the inverted emission of a dipole antenna \cite{Gonoskov.pra.2012,Gonoskov.prl.2013}. This field configuration is also advantageous for the generation of QED cascades, both shower- and avalanche-types, producing copious amounts of high energy gammas \cite{Magnusson.prl.2019}, and serving as a radiation beam dump for high energy particle beams \cite{magnusson2019multiplecolliding}. Many of these applications of the multiple colliding laser pulse configuration rely on another interesting property; radiative particle trapping \cite{Gonoskov_PRL_2014,Esirkepov.pla.2015}.  

\subsection{Radiative trapping}
A multiple colliding laser pulse configuration not only enhances pair production due to the BW process, it also has the interesting property of trapping charged particles near the maxima of its electric field \cite{Gonoskov_PRL_2014}, and occurs only in sufficiently high field intensities. This trapping originates from the general tendency of charged particles to align their motion along the radiation-free direction when they rapidly lose energy and enter the radiation-dominated regime\cite{Gonoskov.pop.2018} and can even be observed in simple mechanical systems with strong friction forces and strong periodic driving \cite{Esirkepov.pla.2017}. 

It has also been recently shown that charged particles in the strong EM fields formed by colliding pulses tend to move along trajectories that are defined by either attractors or limit cycles \cite{Esirkepov.pla.2015,Jirka_PRE_2016,Kirk_JPP2016,Bulanov.jpp.2017}. While such behavior is mostly attributed to intense radiation losses that are typically modeled based on the classical description of radiation reaction, in terms of either a Landau-Lifshitz (LL) equation of motion or a ``modified'' LL equation, the mechanisms behind the observed phenomena are tolerant to the quantized nature of emission and similar behaviour has also been observed in computer simulations based on probabilistic quantum treatment of radiation losses \cite{Jirka_PRE_2016}. 

\subsection{Collective Processes in the QED-Plasma Regime}

Collective plasma processes in the QED-Plasma regime are expected to be dramatically different from the well studied classical plasmas. There are several examples of such processes in the literature already{, starting from radiation reaction effects, to }electron-positron pair production in plasma by plane EM waves \cite{Bulanov_PRE_2004}, which does not happen in vacuum, to the backreaction of pair production on the properties of the EM wave due to the created electron-positron plasma \cite{Bulanov_PRE_2005}, to the laser absorption by created electron-positron plasma during the avalanche-type cascade\cite{Nerush:PRL2011,Grismayer_PoP_2016}, to the laser driven ion \cite{Tamburini_NJP_2010,Del_Sorbo_NJP_2018} and electron \cite{Vranic_PPCF_2018} acceleration, to the reversal of relativistic transparency in QED plasma \cite{Zhang_NJP_2015}. In the last case, electrons in the plasma are accelerated to such high energy in the ultra-relativistic regime that their effective mass is much greater than their rest mass, leading to a reduced plasma frequency by a factor $1/\langle\gamma\rangle$, where $\langle\gamma\rangle$ is the average Lorentz factor of the electrons. Consequently, "relativistically induced" transparency may occur: an opaque (and nominally overdense) plasma may become transmissive, if $\langle\gamma\rangle$ is sufficiently high \cite{Kaw_PF_1970}. However, in the QED-plasma regime, radiation reaction becomes significant, the electron motion is damped and hence $\langle\gamma\rangle$ is reduced. Furthermore, at even greater laser intensities, sufficiently dense pair plasma may be produced to shield the laser fields. Thus, a relativistically transparent plasma would become opaque for laser pulses due to QED effects \cite{Zhang_NJP_2015}. 

\begin{figure}
\begin{center}
    \includegraphics[width=0.48\textwidth]{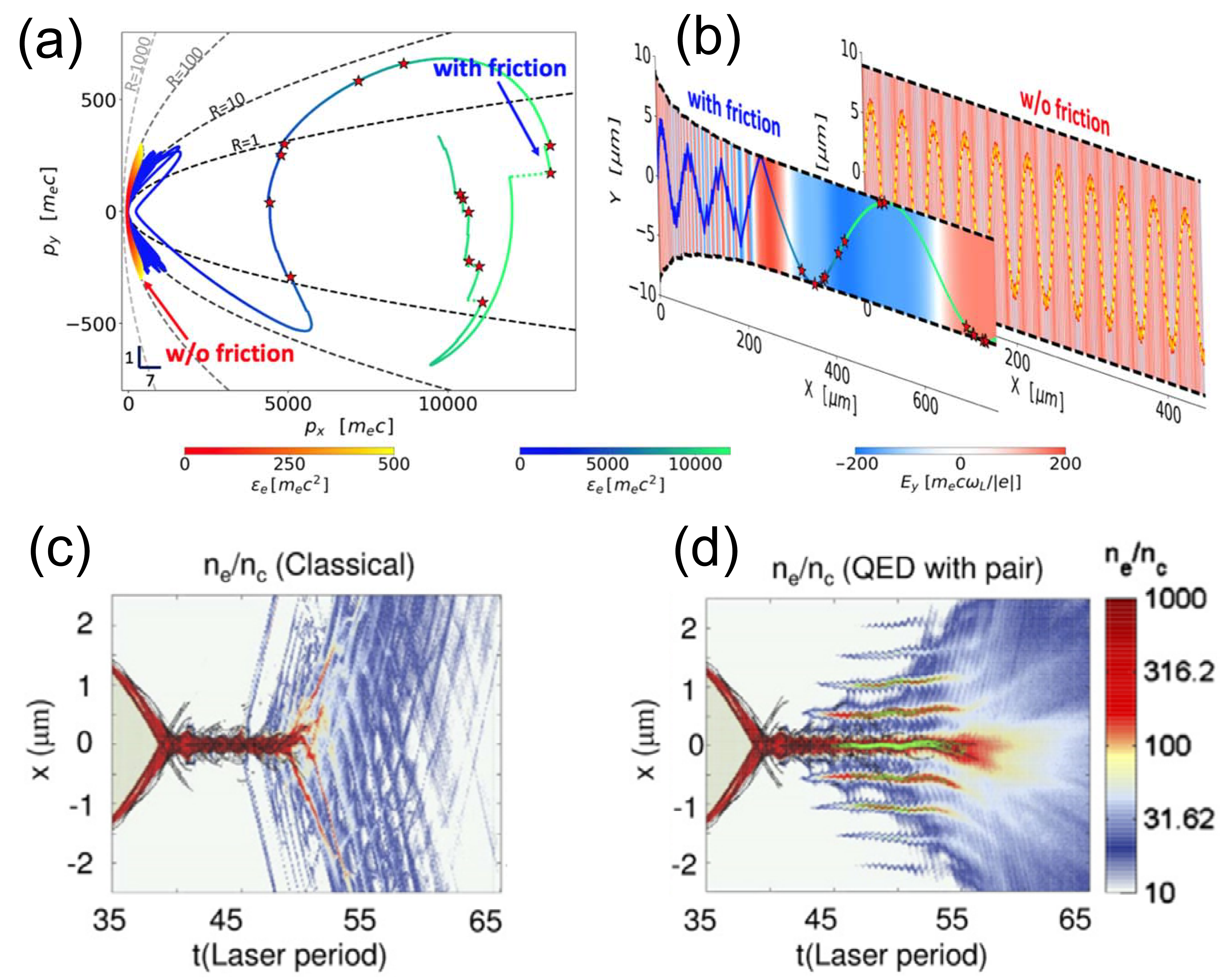}
\caption{(a) Electron trajectories for laser-driven ($a_0 \approx 200$) electron acceleration in a strong magnetic field assisted by radiation friction, in $(p_x,p_y)$ momentum-space without and with radiation friction (instantaneous energy in red-yellow and blue-green colorbar, respectively). The stars indicate photon emission with energy $\varepsilon_{\gamma} > 50$~MeV. (b) Same electron trajectories as in (a) in $(x,y)$ coordinate-space (energy color-coded as in (a)) and the transverse laser electric field at the electron’s position (red-blue colorbar). (c) Electron density $n_e/n_c$ ($n_c$ is the plasma critical density) as a function of time $t$ and position $x$ from 1D QED-PIC simulation for two counter propagating lasers interacting with a foil, when QED is off, and (d) when QED is on. Also plotted are the contours for ion density (black lines) and positron density (green contour lines). (a) and (b) are reproduced with permission from Z. Gong, et al., Sci. Rep. 9, 17181 (2019). Copyright 2019 Nature Publishing. (c) and (d) are reproduced with permission from P. Zhang, C. P. Ridgers,   and A. G. R. Thomas, New J. Phys.17, 043051 (2015). Copyright 2015 IOP Publishing.}
\label{fig:figz1}
\end{center}
\end{figure}

Configurations with a strong collective plasma field~\cite{Stark_PRL_2016} can serve as a novel test-bed for studies of radiation reaction. For example, the radiation reaction or radiation friction is able to enhance the electron energy gain from the laser field even though it is an energy loss mechanism~\cite{Gong_2019}. Figures~\ref{fig:figz1}(a) and (b) illustrate how an electron with a significant transverse momentum becomes strongly accelerated after losing some of its transverse momentum due to radiation friction. The remarkable aspect here is that no energy gain takes place if the radiation friction is not included into the analysis. The red stars in  Fig.~\ref{fig:figz1}(a) show the emission of photons with energy above 50 MeV. It is evident that the emission process is not classical at high electron energies: the photons are not emitted continuously and each of the emissions significantly reduces the electron energy.

QED PIC simulations show radiation reaction dramatically changes the dynamics of QED plasma in the configuration of a thin foil plasma illuminated from two sides by two counterpropagating laser pulses\cite{Zhang_NJP_2015}. When {SF} QED is turned off, hot, back-injected electrons are accelerated away from both sides of the plasma slab (Fig.~\ref{fig:figz1}(c)). When SF 
QED is included, these hot electrons are radiatively cooled such that they get trapped in the nulls of the ponderomotive potential. These electrons form equally spaced ultra-high-density thin electron and positron layers, of approximately equal density, in the nodes of the standing wave formed by the incident and reflected wave (Fig.~\ref{fig:figz1}(d)). 

\subsection{Numerical models for QED plasma}
\begin{figure}
    \includegraphics[width=0.5\textwidth]{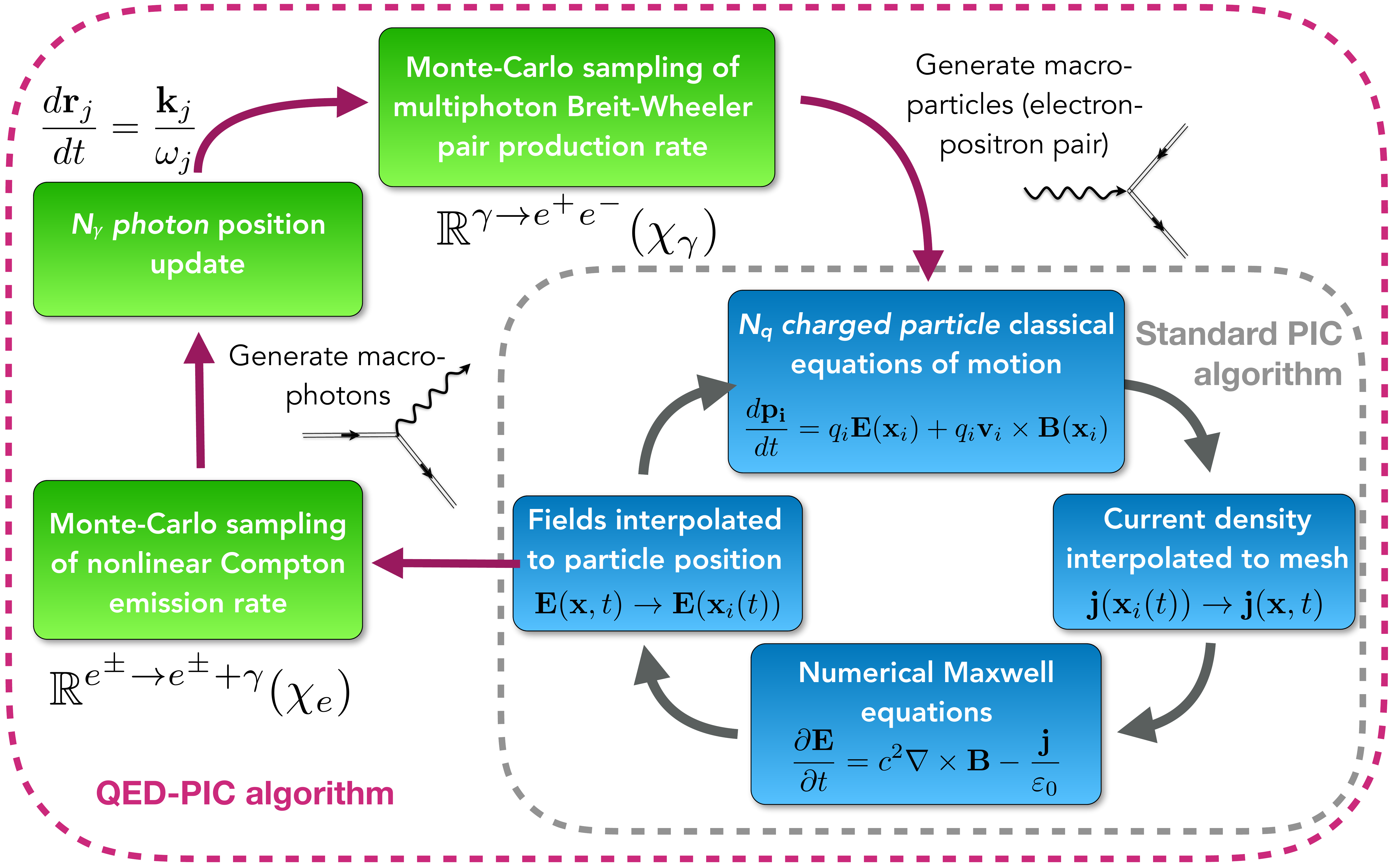}
\caption{Framework of QED PIC. }
\label{fig:figz3}
\end{figure}

The study of QED plasma (as exemplified in the previous subsections) relies extensively on particle-in-cell (PIC) methods with QED extensions. The QED PIC is typically implemented by coupling the QED processes, such as gamma-ray photon emission by electrons and pair production by gamma-ray photons, through a Monte Carlo algorithm to the classical particle-in-cell code \cite{Duclous_PPCF_2011,Ridgers_JCP_2014,Gonoskov_PRE_2015,Vranic:NJP2016}, as illustrated in Fig.~\ref{fig:figz3}. The electromagnetic field is split into high (i.e.~gamma ray) and low (i.e.~optical/plasma) frequency components. The low frequency components are coherent states that are assumed to be unchanged in QED interactions. The evolution of these macroscopic fields is determined by solving Maxwell's equations on the grid. Note that in strong fields Maxwell's equations may be modified by non-linear field-dependent effects \cite{Marklund_RMP_2006}. Electron and positron basis states are influenced by these low-frequency fields, which are treated as a classical background that interacts with the charged particles and the high frequency component of the field, using the strong-field QED representation. The motion of electrons and positrons is subject to the Lorentz force on classical trajectories between point-like QED interaction events. 

{Despite b}eing the ``backbone'' of QED plasma studies, the benchmark of the QED PIC results against experiments is, however, generally missing. With the development of new experimental facilities with extreme laser intensity, experimental validation of the QED PIC calculations is urgently needed. It is also important to understand the limitations of various simplified assumptions made in the QED PIC calculations and when these assumptions break down. The validations against experiments would help build a more robust and accurate QED PIC. 

Alternative approaches to PIC include Boltzmann-like kinetic quantum plasma equations \cite{Neitz_PRA_2014,Ridgers_JPP_2017}, which have the advantages of not having statistical noise and are able to handle particle creation/destruction processes easily. {As a large number of new particles (such as gamma photons and pairs) are created over a short period of time, QED PIC will become computationally costly and time consuming. QED PIC relies on Monte Carlo sampling of the various QED processes (Fig. 6), which will intrinsically introduce statistical noises in the calculation. These shortcomings are not shared by the Boltzmann/Vlasov based approaches, which calculate only the particle energy distribution functions. The latter also has the advantage of handling particles in the high energy tail of the distribution, which are challenging in typical QED PIC.}

\section{\label{sec:level7}Strategies for progress}

\subsection{Theory and Simulations}
There are a number of unanswered fundamental questions in SF QED physics, the better understanding of which would be essential to study the relativistic plasma physics in supercritical fields:

\paragraph{Beyond the plane wave approximation.}

Present-day laser systems achieve very high intensities localized in a small space-time volume with a characteristic size of a wavelength and the electromagnetic field has a complicated, 3-dimensional structure.
However, most studies of the fundamental strong-field QED processes are performed for plane monochromatic waves or constant crossed fields. In these cases, the Dirac equation for an electron in the strong field has exact solutions, which enable one to quite easily obtain analytic formulae for the probabilities of quantum processes. This is possible since plane waves are null fields with high degree of symmetry \cite{Heinzl_PRL2017}. {Most 
}field configurations, especially in multi-laser beam collision scenarios do not possess such {a} high degree of symmetry.
{There have been studies of QED processes for focused fields \cite{Piazza_PRL2014,Piazza_PRL_2016}, but the results apply only in the regime $\gamma \gg a_0$.} 
Finding new (approximate) analytic solutions to the Dirac equation and corresponding probabilities for quantum processes in more realistic field configurations, and the influence of a background plasma medium on them, is important not only for benchmarking results from QED-PIC simulations against them, but also to better understand the limitations of
the QED-PIC approach and identify situations where a self-consistent treatment of these processes in the case of a non-plane wave is required \cite{Raicher_PRA2013,Raicher_PLB2015,King_PRD2016,Mackenroth_PRE2019}.

\paragraph{Beyond the local constant field approximation.} 

 If the formation length/time of a quantum process is much smaller than the respective spatial and temporal inhomogeneities of the laser pulse, the local probability of the process can be calculated in the framework of local constant field approximation (LCFA) model, which is almost always used in present day calculations. However, it is already understood that {the formation length of low energy photons is not small compared to }
 the respective spatial and temporal {inhomogeneities} of the laser pulse. Moreover, as higher intensities are being used, plasma response leads to the generation of strong and very localized EM fields. In this case the formation lengths of the QED processes may become comparable {to} or even longer than the scale of spatial and temporal {inhomogeneities} of these fields.

{For the photon emission process it been discussed discussed that the LCFA fails at low light-cone energy of the emitted photon \cite{Piazza_PRA_2018}.
An} explicit benchmarking of the validity of the LCFA was performed by \citet{Blackburn_PoP_2018}, {by comparing calculations made using LCFA with one using full QED}, especially for the regime $a_0 \lesssim 10$ which is relevant for present day experiments. Corrections and improvements of the LCFA rates have been recently proposed, e.g.~by including local field gradients \cite{Ilderton_PRA_2019} or by explicitly checking whether the radiation formation length is not short \cite{DiPiazza_PRA_2019,Blackburn_beaming}. In either case one needs not only the local field values, but also derivatives along particle trajectories. In addition, the usual assumption of collinear emission has been scrutinized \cite{Blackburn_beaming}. It remains a significant future research program to improve QED-PIC codes by implementing those enhanced rates.

\paragraph{Inclusion of lepton spin.}

With increasing interest in high intensity laser-plasma interactions, it is essential to also understand the interplay between lepton spin effects and the overall plasma dynamics. Not only will the electron spin vectors precess in strong fields, but also the fundamental QED process of photon emission and pair production are all spin-dependent \cite{ST_book,Ivanov_EPJC2004,Ivanov_EPJC2005,Seipt_PRA_2018}, as are radiation reaction effects\cite{DelSorboPPCF}. Consequences of the latter are, for instance, altered equilibrium orbits in the radiation dominated dynamics in rotating electric fields \cite{DelSorboPPCF}, or spin-polarization dependent deflections of electrons in laser electron beam collisions \cite{Geng2019,Li_PRL2019}.

Due to asymmetries in the spin-flip rates electrons can spin-polarize as they interact with high intensity laser pulses, a phenomenon known as Sokolov-Ternov effect from lepton storage rings, where the electron's spins align anti-parallel to the magnetic field. In the magnetic nodes of two counterpropagating circular laser pulses the orbiting electrons can spin-polarize perpendicular to their plane of motion within a few femtoseconds \cite{DelSorboPRA,DelSorboPPCF}. In collisions of electron beams with linearly polarized laser pulses one needs to break the symmetry of the oscillations of the magnetic field in order to establish a distinguished "down" direction.  This can be achieved, e.g., by using ultra-short\cite{Meuren:PRL2011,Seipt_PRA_2018} or bichromatic laser pulses, where both polarized electron and positron beams could be generated\cite{Seipt2019,Chen2019}.

The studies of spin-polarization effects in the context of QED-plasma interactions have only started recently and more work is required. For future progress these spin-interactions need to be included into QED-PIC codes.

\paragraph{Multi-staged processes.} The most straightforward examples of multi-staged processes are avalanche- and shower-type cascades, which are fascinating phenomena of fast transformation of laser and/or charged particle beam energy into high-energy photons and $e^+e^-$ pairs. These processes bring to life a plethora of other effects, such as radiative trapping, attractors and chaos in charged particle motion, and generation of high energy photon and positron sources. A question whether the maximum attainable laser intensity is determined by the cascade development \cite{Fedotov.prl.2010,Bulanov1_PRL_2010} also falls into this category. Theoretical and simulation studies of the cascades and related processes are usually relying on the fact that formation length/time is much smaller than the respective spatial and time inhomogeneities of the electromagnetic field. There is an initial effort in addressing the analytical calculations of the multi-staged processes. Up to now two stage processes were investigated, such as double Compton \cite{Jentschura_PRL,Seipt_PRD_2012,Mackenroth_PRL_2013,King_PRA2015,Dinu_Doublecompton} and Trident \cite{Hu:2010,Ilderton:PRL2011,King_PRD2018,Mackenroth_Trident,Dinu_Trident,HernandezAcosta2019} processes. However, a full QED treatment of multi-staged cascade processes is still to be achieved.

\paragraph{Beyond the external field approximation.} One of the open questions is the back reaction of the processes of either pair production or photon emission on the intense electromagnetic field. Usually these processes are considered using an external field approximation, which assumes that the external field has infinite energy. However, in a number of papers it was pointed out that the creation of new particles can lead to the depletion of the electromagnetic field energy, which invalidates the approximation of the external field \cite{Narozhny_PLA_2004,Bulanov_PRL_2010,Nerush:PRL2011,Seipt_PRL_2017,Ilderton2018,Heinzl:PRD2018}. {For example, it would require $N_e\sim 10^{12}$ of 10 GeV electrons localized in a $\lambda^3$ volume to deplete the EM field with $a_0\sim 10^{3}$, according to the condition $a_0^{1.08}\gamma^{-0.92} N_e\sim 6.8\times 10^{11}$ from Ref. \cite{Seipt_PRL_2017}.} Moreover, in the context of QED-plasma studies, one needs to understand how SF QED processes both backreact on and are initiated by plasma fields. Up to now such studies are performed either on the level of analytical estimates, or by employing PIC QED approach. The ultimate goal for theory here would be a consistent ab-initio real-time description of all quantum-plasma phenomena.  

\paragraph{Breakdown of the quasi-classical approximation in extremely strong fields and the Ritus-Narozhny conjecture.}

The QED-PIC method is based upon a separation of time-scales: That the formation time for quantum processes is very short with classical propagation between incoherent quantum events. However, calculations (for constant crossed fields) show that in extremely strong fields, $\chi \gg1 $, the mean free paths for electrons and photons are on the order of the Compton wavelength $\lambdabar_C \sim 1/m$ for $\alpha\chi^{2/3} \sim 1$. Of course, the concept of a classical particle and thus classical motion has no meaning on the Compton scale, seriously challenging the applicability of QED-PIC at extreme field strengths \cite{Fedotov_2017}.

Calculations of the radiative corrections indicated that those loop corrections in strong-field QED might increase with a power of the energy scale, instead of a logarithmic increase in the absence of strong fields \cite{Ritus_1970,Narozhny_PRD_1980, Fedotov_2017}. This is sometimes referred to as "Ritus-Narozhny conjecture": That for $\alpha \chi^{2/3}\sim 1$ the semi-perturbative expansion of SF QED {(i.e.~perturbative [tree-level] interactions of quantized photons and non-perturbatively laser-dressed fermions)} breaks down and an exact theory of the interaction with the radiation field is required, taking into account all radiative corrections.

However, recent studies\cite{Ilderton_PRD_2019,Podszus_PRD_2019} showed that there seems to be no universal behavior for
$\chi\to \infty$, which is basically a product of field strength and particle energy, when the calculations are performed for short laser pulses instead of constant crossed fields.  The power law scaling $\sim \alpha \chi^{2/3}$ from constant crossed fields was recovered only in a low-frequency-high-intensity limit, and the high-energy limit yields a logarithmic scaling coinciding with ordinary QED. This extreme-field regime is mostly uncharted territory. Further studies are required to resolve the connections and the possible transitions among different power law scalings under various input conditions.

\paragraph{Other theoretical problems}
Other important theoretical problems include the study of the properties of the quantum vacuum under the action of strong fields \cite{DiPiazza_RMP_2012}, radiation corrections, including Cherenkov radiation in strong fields \cite{Macleod_PRL_2019,Bulanov_PRD_2019}, beam-beam interaction in QED plasmas \cite{Yakimenko_PRL_2019}, as well as the manifestations of the physics beyond the Standard Model. Alternative simulation methods not relying on PIC may also be explored, such as real-time lattice QED \cite{Hebenstreit_PRD,Shi_2018}.

\subsection{Experiment and Facilities}

Experiments in this area depend highly on the availability facilities capable of reaching large values of $\chi$, since it will not be feasible in the near term to achieve the critical field strength in the laboratory frame. We note that reaching large values of $\chi$ is an important problem by itself that needs to be addressed by future facility designs. It is due to the fact that electrons and positrons quite easily radiate their energy away when interacting with strong EM fields, so several approaches were proposed to counter this energy loss \cite{Blackburn2019,Baumann:PPCF2019,magnusson2019multiplecolliding} to ensure that high energy particles reach the region of highest field intensity. 

\begin{figure}
    \centering
	\includegraphics[width=\columnwidth]{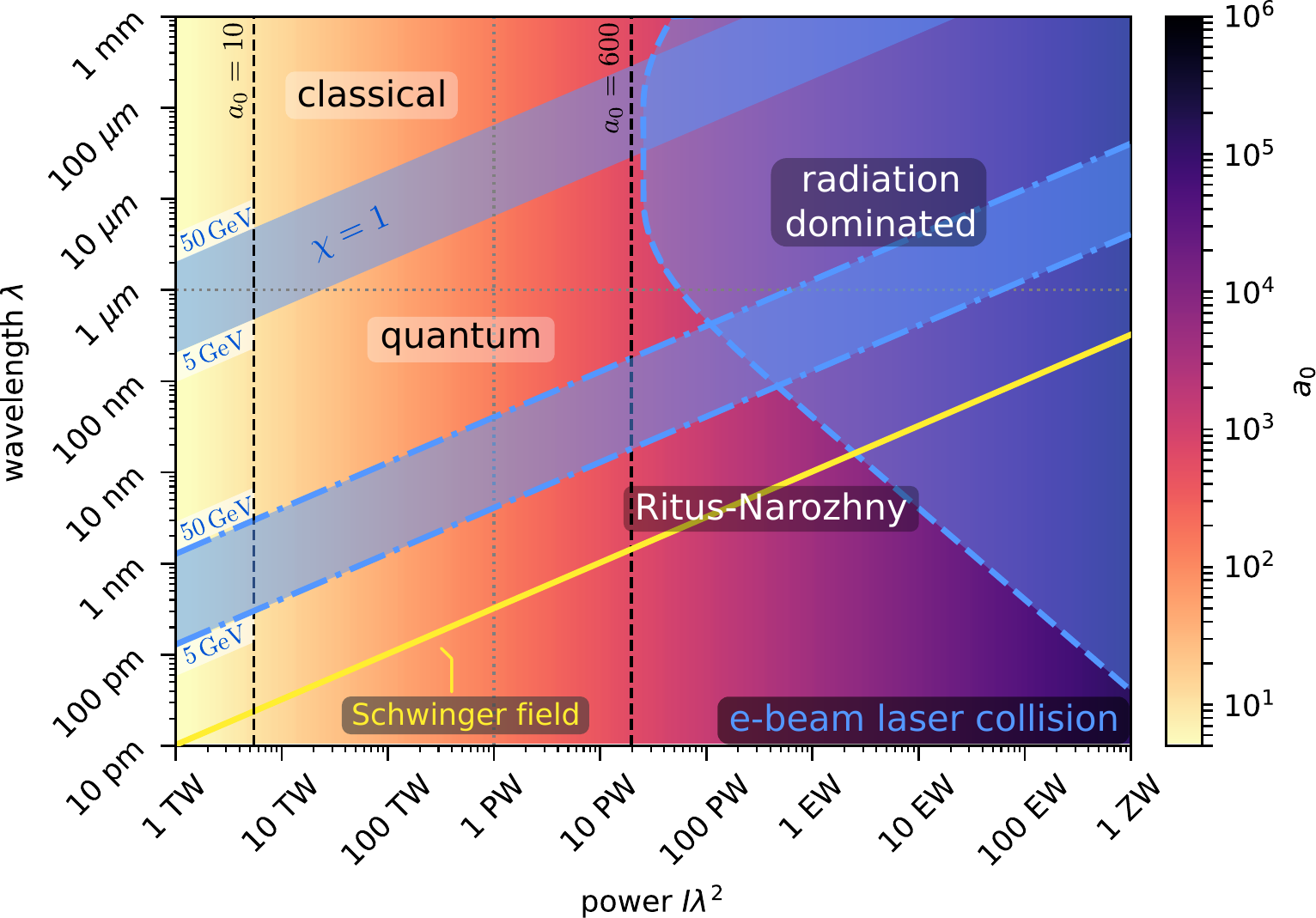}
	\includegraphics[width=\columnwidth]{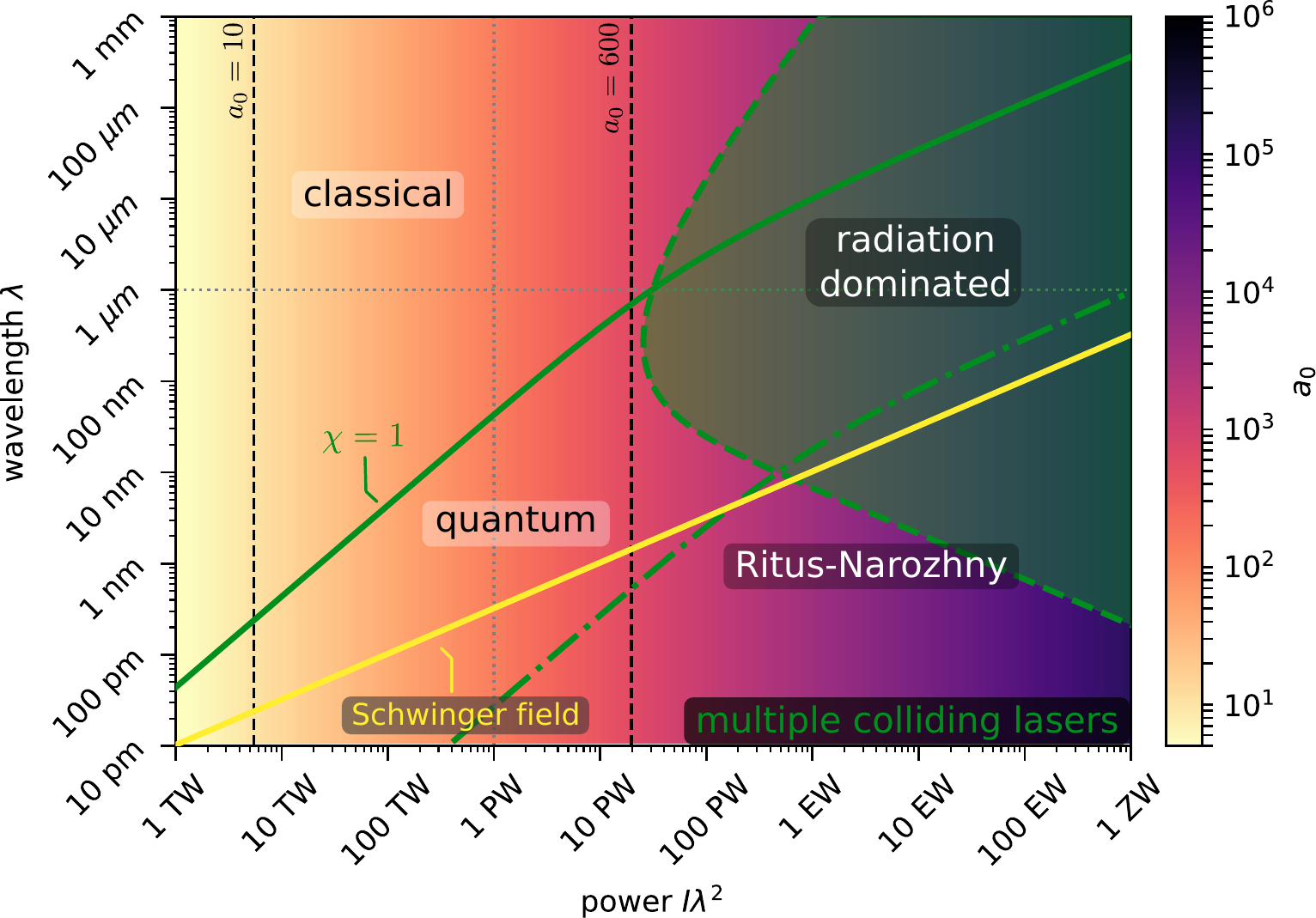}
   \caption{The different regimes of SF QED plasma interactions can be reached with various laser power and wavelength for the e-beam laser collider (upper) and
   the multiple-laser beam interactions (lower).
    The blue diagonal band in the upper plot marks the transition from the classical for the quantum regime at $\chi=1$ for colliding beams of various electron beam energy.
    The green solid curve in the lower plot is the same for the multiple laser interaction. Below the dash-dotted lines is the Ritus-Narozhny regime (Note that it is most easily accessed using short-wavelength radiation in the collider scenario). The shaded regions right of the dashed curves is the radiation dominated regime. The intersection point of the quantum-classical transition and the transition to radiation dominated dynamics occurs around 30 PW and 1 $\mu$m for the multiple laser beam interaction case. 
    We assume focusing to a $2\lambda$ spot size for both plots.
    \label{fig:power_wavelength}
    }

\end{figure}

As has been described in Section II,  critical fields in the zero momentum frame of a high energy particle (pair) can be achieved with two basic configurations: either an externally accelerated relativistic charged particle beam interacting with a perpendicular field or a particle orbiting in a rotating field configuration. In addition to that, the choice of laser wavelength plays an important role in what areas of the interaction parameter space can be accessed \cite{Esirkepov.pla.2015,Bulanov.jpp.2017}. In Figure \ref{fig:power_wavelength} we show how the wavelength of lasers used affects the ability to access strong field QED regimes (defined by reaching $\chi=1$) and strongly radiation dominated regimes [defined by reaching $\alpha a_0 \chi g(\chi) = 1$, where $g(\chi)\leq 1$ describes the reduced radiated power in the quantum regime\cite{Kirk_PPCF_2009,Ritus_1979}], in either a multiple colliding laser configuration (lower panel) or laser colliding with a 5-50GeV lepton beam (upper panel). It is clear from these charts that very short wavelength lasers are able to reach the $\chi=1$ limit most easily and may therefore be the optimal experimental platforms to study nonlinear QED processes. The technology able to produce very short wavelength lasers with required power is not yet developed. Moreover, to study the physics of relativistic \emph{plasmas} in supercritical fields, where the process rates are sufficient that the quantum processes affect the plasma dynamics, we also need to be in the radiation dominated regime. For the multiple laser pulse configuration, the crossing point where radiation dominated and quantum dominated regimes are simultaneously important is near 1~$\mu$m wavelength at 10's of PW laser power. 

\begin{figure*}
\begin{center}
    \includegraphics[width=0.75\textwidth]{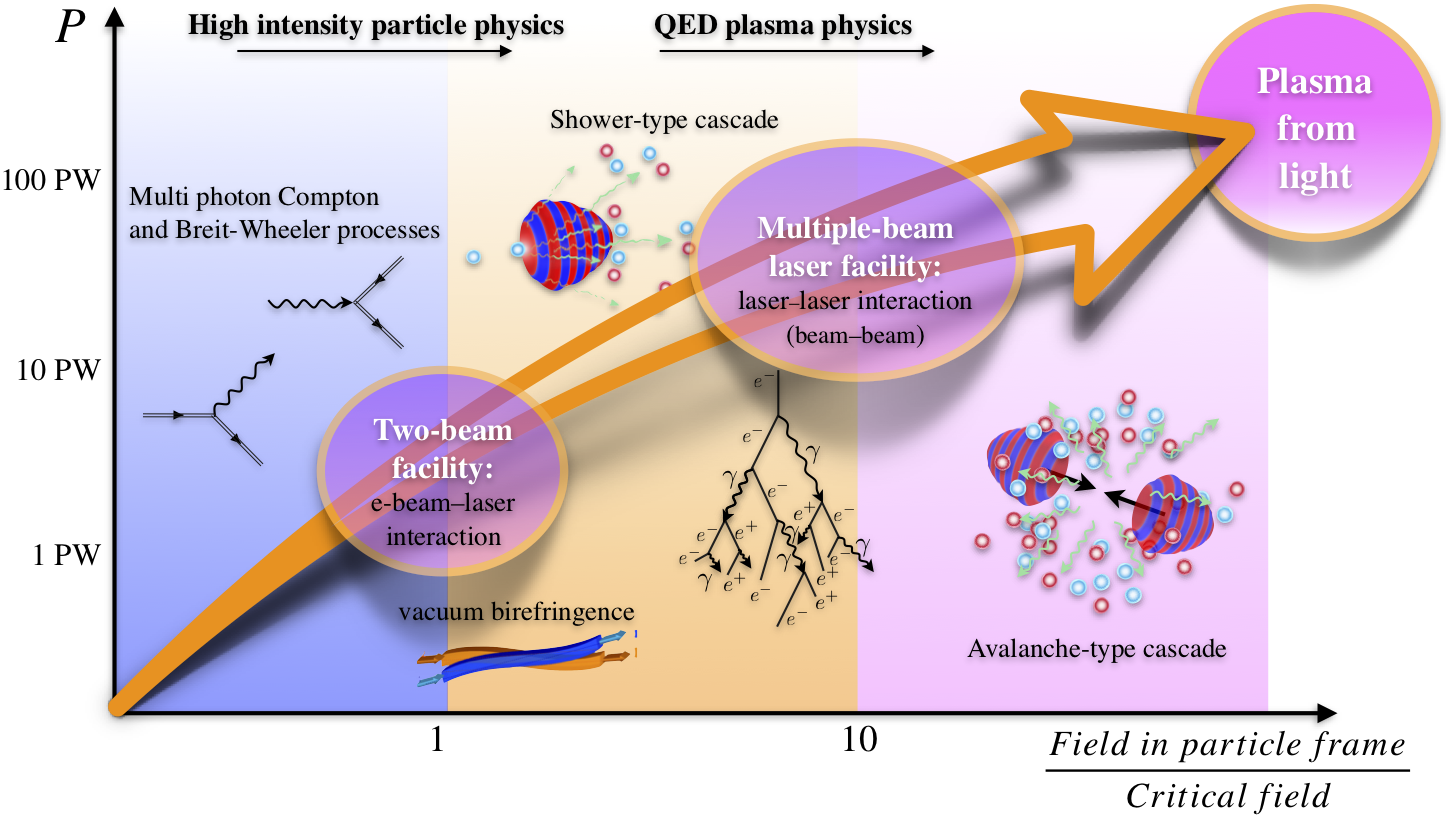}
\caption{Timeline of the QED-plasma studies envisioned as a three-stage process with a facility at intermediate laser intensities for the study of fundamental strong-field QED processes, a multi-beam facility at high laser intensities to study the interplay between collective plasma effects and strong-field quantum  processes, and a facility based on laser-plasma collider to study the ultimate limits of SF QED. Reproduced with permission from 
H. M. Milchberg and E. E. Scime,  “Workshop report: Workshop on opportunities, challenges, and best practices for basic plasma science user facilities (May 20-21, 2019, University of Maryland, College Park, MD),” arXiv:1910.09084.
}
\label{fig:fig3}
\end{center}
\end{figure*}

We  therefore envision three basic stages of high-power laser facility development for experimental research into QED-plasma regime, as illustrated in Fig.~\ref{fig:fig3}: (i) PW-class laser facility with an additional colliding beam, (ii) multibeam laser facility with 10-100 PW of total power, and (iii) laser plasma collider, featuring multiple PW-class lasers, with 0.1-1EW of total power, capable of being reconfigured into {either an} e-beam laser {collider} {or }
multiple beams setups. The following discussion is mostly aimed at $\sim$1 $\mu$m lasers which are the most prevalent and technologically advanced ultra-high peak power laser technology today.
{An alternative is beam-beam interactions, see Ref\cite{Yakimenko_PRL_2019} for further details.}

\paragraph*{Stage 1 (facility)}: The study of basic quantum processes of strong field QED in the high-intensity particle-physics regime together with the relativistic plasma physics phenomena can be carried out at a PW-class laser facility featuring an additional colliding beam. This could mean either with two laser beamlines, with one of them being used for particle acceleration, or with a laser and an external electron beam.  The main laser beamline with power $P_\mathrm{laser}$ should be focusable to a spot-size of order a wavelength $\lambda_\mathrm{laser}$  such that the product $\left(\mathcal E_\mathrm{beam}\;{\rm[GeV]}\right) \sqrt{P_\mathrm{laser}\;{\rm[PW]}}\left(\lambda_\mathrm{laser}\;{\rm[\mu m]}\right)^{-1}\gg1$, where $\mathcal E_\mathrm{beam}$ is the beam energy. Apart from the stability and high repetition rate\cite{Arran_PPCF_2019}, such facility should be able to provide the parameters of interaction that would allow the experimental mapping of the transition from the classical to quantum description of the interaction. This means that the parameter $\chi$ should vary from 0.1 to 10, {\it i.e.,} $2\times 10^5<\gamma a_0 <2\times 10^6$. These parameters do not look extreme for existing laser facilities, since the current record for laser wakefield acceleration (LWFA) electron beam is $\sim 8$ GeV (or $\gamma\approx 1.6\times 10^4$) \cite{Gonsalves_PRL_2019}, and the peak laser intensity achieved so far is ${5.5}
\times 10^{22}$W/cm$^2$ {\cite{Yoon_OP_19,Yanovsky:08,Kiriyama:18}}, which is $a_0\sim {130}
$. Thus, these parameters can be achieved by either employing a PW laser and a 10 GeV electron beam, or a few GeV electron beam and a multi-PW laser pulse. 

Though this setup is similar to that used in E144 experiment at SLAC, and in two recent ones at GEMINI, it is of critical importance to study SF QED effects in such a configuration with higher laser intensity, so that the threshold $\chi=1$ can be exceeded for the first time, and with significantly higher precision and higher statistics. Thus, such facility should be capable of producing high-energy high-intensity, stable, and high repetition rate collisions between an electron beam and a laser pulse, allowing for the experiments with statistical significance, i.e, "science with error bars". 

\paragraph*{Stage 1 (Experiment)}: The experiments at such facility will be aimed at the study of electron beam collision with an intense laser pulse. Thus, the facility will be able to provide insight in a number of important SF QED problems. It would provide a testbed for PIC QED codes and analytical calculations, testing the {\it plane wave} and {\it local constant crossed field} approximations. It will probe the interplay between {\it lepton spin effects} and the overall plasma dynamics. At high values of $\chi$ these facilities will generate {\it multi-staged processes}, where Compton and Breit-Wheeler processes will follow each other in quick succession multiple times, {\it i.e.} shower type cascades. Moreover, these facilities may map a way towards new high  brightness sources of high energy photons and positrons, which can be utilized for different applications in fundamental physics and material science. In addition, laser ion acceleration experiments can be conducted at such facilities, with laser intensities high enough to probe the onset of SF QED effects during the interaction. This would help to map the transition from the relativistic plasma physics domain into the QED-plasma domain. These facilities will mainly operate in the 
{single particle relativistic electrodynamics, high-intensity particle physics, and relativistic plasma physics parameter space, see also Figure \ref{fig:fig1}.}

\paragraph*{Stage 2 (facility)}: In order to access the QED-plasma regime through either avalanche-type cascades or interactions with different plasma targets and probe the transition of the interaction from particle dominated to radiation dominated (see Figure \ref{fig:power_wavelength}), a facility capable of delivering multiple laser pulses to the interaction point at extreme intensities is needed. Assuming focusing to a spot-size of order of a wavelength, the laser power should satisfy $P_\mathrm{laser}\;{\rm[PW]}\left(\lambda_\mathrm{laser}\;{\rm[\mu m]}\right)^{-1}\gg10$ to achieve that, which brings the total facility laser power into 10's to 100's of PW domain. An alternative configuration for reaching these conditions could involve two extremely high energy and tightly focused lepton beams in a collider configuration \cite{Yakimenko_PRL_2019}, which {has the advantage of being able to reach high $\chi$}, however, will be limited to the study of cascades and beamsstrahlung. 

\paragraph*{Stage 2 (experiment)}: The multiple-beam facilities will mainly be aimed at the study of the {\it multi-staged processes} and a plethora of phenomena that follow them. This includes avalanche-type cascades, radiative trapping, and the transition from chaotic to regular motion for charged particles. As the theory for these processes is being developed and the numerical tools are being correspondingly upgraded, the stage 2 facilities will play a key role in verifying theoretical and simulation results. We envision several types of experiments studying avalanche-type cascades, prolific production of electrons, positrons, and high energy photons, the interaction of this emerging $e^+ e^- \gamma$ plasma with multiple laser beam configuration. These cascades can be seeded by either an initial plasma target, or an external particle or photon beam, and can lead to the generation of high energy high brightness photon source for different applications. The focus of the experiments will not only be the observation of the corresponding processes, but the understanding how SF QED processes both back-react and are initiated by plasma fields. Here the {\it external field approximation} will be tested and the limits of it will be determined. These facilities will also be well suited to study the laser ion acceleration to relativistic {ion energies}, which will be heavily influenced by QED-plasma effects. 
{These facilities will operate in the single particle relativistic electrodynamics, high-intensity particle physics, relativistic plasma physics, and QED-plasma parameter space.}

\begin{figure*}
\begin{center}
    \includegraphics[width=1\textwidth]{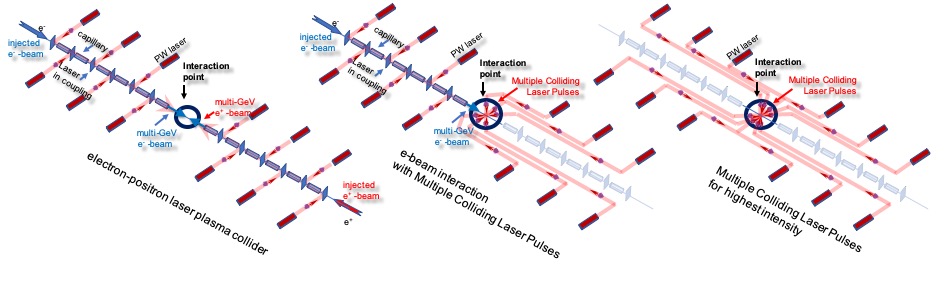}
\caption{The principle scheme of the ultimate SF QED facility, housing multiple PW-class lasers. Depending on the designed c.m. energy the number of stages can be increased correspondingly. The facility can operate in several modes, including (i) $e^+e^-$ laser collider with all lasers utilized to drive the staged acceleration of electron and positron beams; (ii) electron beam interaction with high intensity EM field, where half of the lasers is driving the staged acceleration of the electron beam, and another half provides high intensity field through the multiple colliding pulses configuration; and (iii) all the laser pulses are brought to the interaction point to generate highest intensity possible through the multiple colliding pulses configuration.}
\label{fig:future facility}
\end{center}
\end{figure*}

\paragraph*{Stage 3 (Ultimate SF QED facility)}: 
An important scientific application of high power lasers is particle acceleration for high energy (density) physics, material science, and bio-medical applications, such as the laser plasma collider \cite{Leemans_PhysToday_2009,Schroeder-PRSTAB10,doe.AAC.report}. This collider is proposed to consist of two LWFA arms, one for electrons, the other for positrons, powered by multiple laser pulses, each responsible for the acceleration of electrons or positrons in its own module \cite{Leemans_PhysToday_2009}. The total power required for a collider is approximately 1 PW per 10 GeV in a single module, which sums up to 10's of PW for a 100's GeV class linac or to 100's PW for a TeV-class machine. We argue that it is natural for SF QED studies to be conducted at the same location with minimal adjustments to the facility configuration. In Fig.~\ref{fig:future facility} we sketch a principal design for SF QED/plasma accelerator facility that would provide an ultimate test to the advanced accelerator technologies as well as to supercritical field effects in high energy physics and plasma physics. Such facility would combine both Stage 1 and Stage 2 capabilities at higher energy and intensity levels. 

When both arms of the accelerator are powered (left pane of Fig.~\ref{fig:future facility}), it will be able to operate as a  high energy physics machine to study electron-positron collisions at the 0.1-10 TeV level. This mode of operation can, in principle, be used to the benefit of SF QED through studies of beamstrahlung, cascades, and beam disruption, in the regime where the breakdown of semi-perturbative expansion of SF QED may occur. 

Another mode of operation will be when only one arm of the accelerator is powered to produce a high energy electron or positron beam (central pane of Fig.~\ref{fig:future facility}). The lasers from the other arm will be rerouted to the interaction point in a form of multiple colliding laser pulses, which would provide a configuration of Stage 1 for the study of different regimes of SF QED in e-beam laser interactions, but with the e-beam energy and EM field strength many times higher than what can be produced by a single laser at Stage 1. 

The third mode of operation will bring all the lasers to the interaction point providing the highest intensity for experiments involving different fixed plasma targets and quantum vacuum properties studies (right pane of Fig.~\ref{fig:future facility}). 

\paragraph*{Stage 3 (experiment)}: The experiments at such facility will address the frontier of SF QED and plasma physics in beam-beam, e-beam laser, and multiple laser collisions in vacuum and using a variety of plasma targets. Since the parameters of operation will be well into the radiation dominated region, which is the domain of QED-plasma physics, the backreaction and initiation of SF QED processes by laser and plasma fields will dominate the interaction. The experiments at this facility will be able to study the behavior of plasma trapped in strong fields. Recent simulations results hint at the unconventional behavior dominated by the effects of SF QED. Such facility will also be well equipped for the study of the effects of beam disruption, beamstrahlung, and final focussing effects in intense beam-beam collisions relevant to collider applications.

The study and verification of quantum plasma theory will be the main experimental goal. 
{We envision a series of experiments at the ultimate SF QED facility addressing the above mentioned breakdown of Ritus-Narozhny conjecture (Sec. IV A 6), since different modes of operation enable experiments not only in both regimes in the high intensity limit ($\chi\rightarrow\infty$, for $E\rightarrow \infty$ and fixed $\gamma$) and in the the high energy limit ($\chi\rightarrow\infty$, for $\gamma\rightarrow \infty$ and fixed $E$), but also in the beam-beam interaction regime.} Furthermore, this facility will make it possible to study the properties of the quantum vacuum: from polarization to "breakdown", as well as to search for the physics beyond the Standard Model.

\paragraph{General experimental considerations}
There are numerous  issues to address to realize these experiments in practice. Having colliding laser pulses or lasers interacting at such high power with anything that may provide back-reflections can send laser energy back along the laser chain and damage or destroy elements. Providing isolation for the laser chain is something that needs development. Overlapping micron scale, femtosecond duration beams requires not only precision with overlap but also a high degree of stability in time and laser beam pointing. High repetition rate is also desirable for collecting statistically significant data and scanning large parameter spaces. High peak power, high average power laser systems are being developed that can push repetition rates up to kHz or beyond, but the state of the art for PW class laser systems is of order a few Hz at present. {An additional experimental challenge is the development of detectors or detection methods for GeV scale high-energy particles.} The development of these novel technologies will greatly benefit future applications in fundamental physics, industry, bio-medical research, and homeland security.

\section{Conclusions}

The new plasma state that is created in the presence of supercritical fields is similar to that thought to exist in extreme astrophysical environments including the magnetospheres of pulsars and active black holes. Electron-positron plasmas are a prominent feature of the winds from pulsars \cite{Goldreich_PRL_1969} and black holes \cite{Ruffini_PR_2010}. 

In terrestrial laboratories, the sources of the highest intensity EM fields are lasers, with the exception of {aligned crystals\cite{Wistisen_NatCom2018,Uggerhoj_RMP2005} and }highly charged ion interactions, the latter
are, unfortunately, dominated by quantum chromodynamics effects. The interaction of lasers with electrons, positrons, and photons,  whether they act as single particles or plasma constituents, may lead to a number of SF QED effects including vacuum "breakdown" and polarization, light by light scattering, vacuum birefringence, 4-wave mixing, high harmonics generation from vacuum, and EM cascades of different types. It was shown theoretically and in computer simulations that one can expect the generation of dense electron-positron plasma from near vacuum, complete laser absorption, or a stopping of an ultrarelativistic particle beam by a laser light.

All these phenomena are of fundamental interest for  quantum field theory. Moreover, they should dominate the next generation of laser-matter interaction experiments, and may be important for future TeV-class lepton colliders. 
Applications  resulting from high-intensity laser-matter interactions, including high energy ion, electron, positron, and photon sources for fundamental physics studies, medical radiotherapy and next generation radiography for homeland security and industry will benefit from advances in this area. 

However, despite the tremendous progress achieved in SF QED theory, computer simulations, and experiment, with and without plasma, there are a number of unanswered questions and topics, which should dominate the attention of the people working in this field for the next decade. From our point of view, serious progress in SF QED theory and computer modeling is not possible without addressing a number of questions connected with widely accepted approximations. Future studies need to go beyond (i) the plane wave approximation, (ii) the local constant field approximation, and (iii) the external field approximation. All these approximations permit analytical treatment of SF QED processes in the framework of semi-perturbative theory and are straightforward to implement in computer modeling. But they oversimplify the structure, the behavior, and the interaction of charged particles and photons with EM fields.As the laser intensity and the strength of plasma EM fields increase, the interaction may enter a regime where semi-perturbative expansion of SF QED breaks down and an exact theory of the interaction with radiation field is required.

Even well before this extreme-field regime in field strength a serious development in SF QED theory is required. It is connected with the fact that at high intensities and high energies the interaction becomes multi-staged. This means that charged particle interactions with these fields lead to multiple photon emissions, which can be accompanied by re-acceleration in these fields.  
 The emitted photons in these fields can decay into electron-positron pairs, which will also start to emit photons multiple times. A full QED treatment of multi-staged cascade processes is still to be achieved. The ultimate goal for theory would be a consistent ab-initio real-time description of all quantum-plasma phenomena.

Basing on the current understanding of SF QED phenomena and recent experimental results, we suggested a staged approach to future experimental studies. Since the experiments depend crucially on the availability of facilities capable of reaching high values of $\chi$, we envision three types of such facilities: (i) PW-class laser facility featuring a second colliding beam, (ii) 10's of PW to EW laser facility able to deliver multiple pulses to the interaction spot to maximize the intensity, and (iii)the ultimate SF QED facility combining the capabilities of the first two with the plasma based lepton collider. These interaction setups correspond to two basic configurations for the study of SF QED phenomena that maximize the parameter $\chi$, or, in other words, 
maximize the EM field strength in the particle rest frame. While these processes will be explored, a number of approximations, used in SF QED theory will be tested. {These} include plane wave, local constant field, and external field approximations. At the highest intensities and the highest energies the validity of the semi-perturbative SF QED expansion will be verified, as well as the phenomena important for the operation of a TeV scale laser driven lepton collider.  

The experiments that were carried out (E144 at SLAC and two recent ones at GEMINI) and are being planned (E320 at SLAC and LUXE \citet{Abramowicz2019} at DESY) are all employing laser e-beam collision to study Compton and Breit-Wheeler processes and their effect on the e-beam behavior. Future experiments should go far beyond that. First, mapping of the transition from classical to quantum description of the interaction is needed. Second, observing multi-staged processes, i.e., cascades (shower- and avalanche-types), should be achieved. Third, the possibility of generating a source of high energy photons and/or positrons should be explored. Fourth, the effects of SF QED in ion acceleration should be identified and explored, as well as the {behavior} of plasma in supercritical fields in the radiation dominated regime. Fifth, the properties of quantum vacuum, including polarization and breakdown, will be studied.   

The study of relativistic plasmas in supercritical fields would help better understanding of many other astrophysical events, such as Gamma Ray Bursts, gravitational collapse, {and} active galactic nuclei.
With the upcoming PW level lasers, soon in the laboratory, we will have access to the conditions in some of the most energetic events in the universe. 

In the US, there are several PW-class laser facilities in operation, but these represent only a fraction of the number and individual power of such facilities being built and in operation around the world. The first experiments to study SF QED effects were performed in the US at SLAC in the 1990s,  but since then the US has lost its leadership in this field both in theory/simulations and in experiment, as indicated in the National Academies of Sciences, Engineering, and Medicine report \cite{NAS_report}. With the rapid development of laser technologies and funding priorities shifted towards high power, high intensity laser facilities, Europe and Asia are leading the field of high-intensity laser-matter interactions, including the \$B Extreme Light Infrastructure in Europe \cite{eli}. In the US, recent developments include the formation of the LaserNet US \cite{lasernetus} community of PW-class lasers, which is increasing collaboration in related areas and the funding of the Zetawatt Equivalent Ultrashort pulse laser System (ZEUS) by the National Science Foundation. This is a multibeam laser system which is designed to address the stage 1 experiments described in this document, by accelerating a multi-GeV class electron beam by laser wakefield acceleration \cite{Gonsalves_PRL_2019} and colliding it with a PetaWatt laser such that the laser power in the electron rest frame is a ZetaWatt. Another SF QED facility proposed in US is EP OPAL at the Laboratory for Laser Energetics \cite{Bromage_2019}, which is envisioned as a multibeam facility with a first stage of development featuring two colliding 30 PW laser beams with the possibility of adding more colliding pairs of beams later. Such facility will be able to address the stage 2 experiments, described above.

\begin{acknowledgments}
{The authors are extremely grateful for useful general discussions and inputs towards the development of figure 8 with the members of the working group on “Plasma in Supercritical Fields” at the 2019 NSF Workshop in Maryland on Basic Plasma Facilities; Gerald Dunne, Sebastian Meuren, Matthias Fuchs, Stuart Mangles, Marija Vranic, Hans Rinderknecht,  and Arkady Gonoskov.} P. Zhang was supported by the Air Force Office of Scientific Research (AFOSR) YIP Award No. FA9550-18-1-0061. S. S. Bulanov acknowledges support from the US DOE Office of Science Offices of HEP and FES (through LaserNetUS), under Contract No. DE-AC02-05CH11231. A.G.R. Thomas acknowledges support from the US DOE Office of Science FES (through LaserNetUS) under grant DE-SC0019255 and the National Science Foundation ZEUS midscale facility under grant 1935950. A. Arefiev was supported by NSF (Grants No. 1632777 and 1821944) and AFOSR (Grant No. FA9550-17-1-0382). 
\end{acknowledgments}


\bibliography{refs}

\end{document}